\documentclass[twocolumn]{autart}    

\usepackage{graphicx}          
\usepackage{amsmath}
\usepackage{amssymb}
\usepackage[noadjust]{cite}
\usepackage{enumitem}
\usepackage{xcolor}
\usepackage{wrapfig}
\usepackage{bbm}
\usepackage{mathtools}
\input{insbox}

 \usepackage{natbib}

\usepackage{siunitx}

\usepackage{hyperref}

\hypersetup{
    colorlinks = false,
    linkbordercolor = {white}
}

\newcommand{\ie}{i\/.\/e\/.,\/~}
\newcommand{\eg}{e\/.\/g\/.,\/~}
\newcommand{\cf}{cf\/.\/~}
\newcommand{\fig}{Fig\/.\/~}
\newcommand{\eq}{Eq\/.\/~}
\newcommand{\sect}{Sec\/.\/~}

\newcommand{\myqed}{$\hfill\qed$}

\DeclareMathOperator{\stcts}{\tau}%
\newcommand{\stdisc}{\tau^\mathrm{d}}%
\newcommand{\stkf}{\tau^\mathrm{o}}%
\newcommand{\vard}{Q}
\DeclareMathOperator{\varc}{\mathcal{Q}}

\usepackage[labelfont=bf,justification=raggedright,singlelinecheck=false]{caption}
\begin{document}

\begin{frontmatter}

\title{Event-triggered Learning} 

\thanks[]{This work was supported in part by the German Research Foundation (DFG)
Priority Program 1914 (grant TR 1433/1-1), the International Max Planck Research School for Intelligent Systems (IMPRS-IS), the Cyber Valley Initiative, and the Max Planck Society.%
}
\author[]{Friedrich Solowjow}\ead{solowjow@is.mpg.de}, 
\author[]{Sebastian Trimpe}\ead{trimpe@is.mpg.de}               
\address{Intelligent Control Systems Group, Max Planck Institute for Intelligent Systems, Stuttgart, Germany}  


\begin{abstract}          
The efficient exchange of information is an essential aspect of intelligent collective behavior.
Event-triggered control and estimation achieve some efficiency by replacing continuous data exchange between agents with intermittent, or event-triggered communication.  Typically, model-based predictions are used at times of no data transmission, and updates are sent only when the prediction error grows too large.
The effectiveness in reducing communication thus strongly depends on the quality of the prediction model.
In this article, we propose event-triggered learning as a novel concept to reduce communication even further and to also adapt to changing dynamics.
By monitoring the actual communication rate and comparing it to the one that is induced by the model, we detect a mismatch between model and reality and trigger model learning when needed.
Specifically, for linear Gaussian dynamics, we derive different classes of learning triggers solely based on a statistical analysis of inter-communication times and formally prove their effectiveness with the aid of concentration inequalities.
\end{abstract}

\end{frontmatter}
\section{Introduction}
Modern communication technology allows for connecting many devices and systems in unprecedented ways.  Thus, enabling applications such as mobile sensor networks, distributed robotics, and multi-vehicle systems, often subsumed as networked control systems (NCSs)~\citep{HespanhaNaghshtabriziXuJan07}.    
Alongside the potential of NCSs come significant challenges for control design such as delayed transmission, packet drops, or limited bandwidth, which  originate from the fact that a shared network is used for feedback.   Many methods that have been proposed for addressing these challenges rely on accurate dynamics models.  For instance, model-based predictions in event-triggered state estimation and control~\citep{lemmon2010event, heemels2012introduction, shi2015event, miskowicz2015event} can replace periodic communication up to a certain extent.  Only if predictions become inaccurate, communication of sensor measurements is indispensable and thus 
\emph{triggered when necessary}. 
In this article, we propose to extend the paradigm of event triggering to model learning and introduce the novel idea of \emph{event-triggered learning} (ETL). 
With ETL, we detect a mismatch between model and true dynamics and trigger identification of a new model whenever needed.
We built ETL on top of a typical event-triggered state estimation architecture (see \fig \ref{fig:ETL-fullstate}) and show that the new architecture can cope with changing dynamics and yields further communication savings in an NCS.

The ability to learn is thus, a fundamental aspect of future autonomous systems that are facing uncertain and changing environments. However, the process of learning a new model or behavior typically  does not come for free but involves a certain cost. For example, gathering informative data can be challenging due to physical constraints, or updating models can require extensive computation. Moreover, learning for autonomous agents often requires exploring new behavior and thus, typically means deviating from nominal or desired behavior. Hence, the question \emph{when to learn} is essential for the efficient operation of autonomous systems. We address this question with ETL, and this article develops the concept specifically in the context of NCSs.

\subsection*{Main Idea: Event-triggered Learning}
\begin{figure}[t]
	\centering
		\includegraphics[width=\columnwidth]{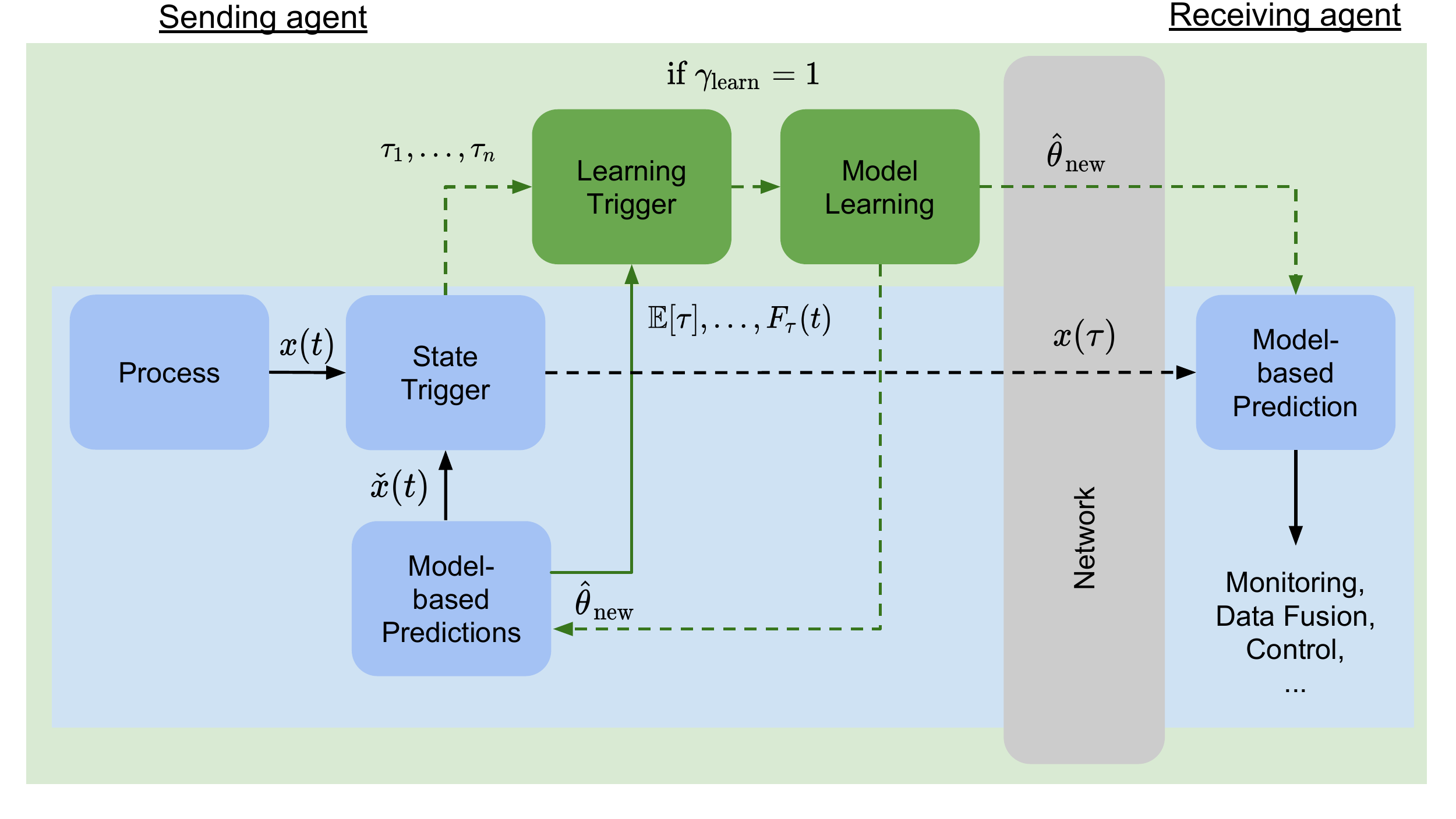}
		\caption{\scriptsize Proposed event-triggered learning architecture for a networked control problem between two agents. Based on a typical event-triggered state estimation architecture (in blue), we propose event-triggered learning (in green) to improve predictions and lower communication between Sending and Receiving agents. Learning experiments are themselves triggered as necessary by comparing empirical with expected inter-communication times.}
		\label{fig:ETL-fullstate}
\end{figure}

We explain the main idea of event-triggered learning using the schematic in \fig \ref{fig:ETL-fullstate}. 
The figure depicts a canonical problem, where one agent (`Sending agent' on the left) has information that is relevant for another agent at a different location (`Receiving agent'). For instance, this setting is representative of remote monitoring scenarios, distributed sensor fusion, or two agents of a multi-agent network. For resource-efficient communication, a standard event-triggered state estimation (ETSE) architecture
is used (shown in blue). The main contribution of this work is to incorporate learning into the ETSE architecture. By designing an event trigger also for model learning (in green), learning tasks are performed only when necessary.
Next, we explain the core components of the proposed framework.

The sending agent in 
\fig \ref{fig:ETL-fullstate}
monitors the state of a dynamic process (either
directly measured or obtained via state estimation) and can transmit this state
information to the remote agent. The true parameters $\theta$ of the process are unknown to both agents.
An event-triggered protocol is used to save network resources.
The receiving agent uses a model (with parameters $\hat{\theta}$) of the process for predicting 
the state at times of no communication. 
The sending agent implements a copy of the same prediction
and compares it to the current state in the `State Trigger', which triggers a state communication whenever the prediction deviates too much from the actual state.
This general scheme is standard in ETSE literature (see~\citep{shi2015event,trimpeEBCCSP15,tr17} and references therein). 
The effectiveness of this scheme will generally depend on the accuracy of the prediction, and, thus, the quality of the model $\hat{\theta}$.

The key idea of this work is to trigger model learning when communication rates deviate significantly from what is expected. 
Because performing a learning task is costly itself (\eg involving computation and communication resources, as well as possibly causing deviation from the control objective), we propose event-triggering rules also for model learning (`Learning Trigger').
Newly learned models are then shared with the remote agent to improve its predictions. 
Since communication itself is triggered by model-based state predictions, we obtain a tractable feature to quantify model accuracy by analyzing the communication pattern. 
Further, we avoid analyzing raw output data, which is possibly multidimensional and highly correlated.  
Thus, we propose a method to obtain improved models from data when needed, which leads to superior communication rates.

While the idea of using event triggering to save communication in estimation or control is quite common by now, this work proposes event triggering also on a higher level. Triggering of learning tasks yields improved prediction models, which are the basis for ETSE at the lower level.

\subsection*{Related Work}
Various event-triggered control~\citep{lemmon2010event, heemels2012introduction, miskowicz2015event} and state estimation~\citep{lemmon2010event, shi2015event, trimpeEBCCSP15, tr17} algorithms have been proposed for improving resource usage in NCSs. 
Approaches differ, among others, in the type of models that are used for predictions.  
The send-on-delta protocol~\citep{miskowicz2006send} triggers data transmission when the difference between the current and last communicated value passes a threshold.
This protocol is extended to linear predictions in~\citep{Su07}, which are obtained by approximating the signal derivative from data.  More elaborate protocols use dynamics models of the observed process, which typically leads to more effective triggering~\citep{trimpe2011experimental, SiKeNo14, trimpe2014event, trimpe2019resource, sijs2012event, wu2013event, battistelli2018distributed,han2015stochastic}. 

Recent articles proposed to improve and augment typical event-triggered state estimation~\citep{shi2014event,battistelli2018distributed,huang2017energy} and control algorithms~\citep{vamvoudakis2018model,vamvoudakis2019robust,baumann2018deep, narayanan2017event} with data-based techniques. 
In these works, learning is used to approximate intractable conditional probability densities that arise in distributed problems or to obtain tractable solutions to Hamilton-Jacobi-Bellman equations that yield optimal control policies, \eg with model-free methods such as Q-learning, or based on neural networks based. 
However, none of these works considers principled decision making on model learning in order to improve prediction accuracy, as we do herein.
Thus, we address the fundamentally different question of \emph{when to learn}, leading to the new concept of ETL. 

In order to obtain effective learning triggers, we take a probabilistic view on inter-communication times (\ie the time between two communication events) and trigger learning experiments whenever the expected communication differs from the empirical.  A similar probabilistic interpretation of inter-communication times is considered in~\citep{XuHe04}, where NCSs are modeled as jump-diffusion processes, and the expected value of inter-communication times is considered. In this article, we analyze the statistical properties in a more general context and propose the design of learning triggers based on inter-communication times and concentration inequalities.

Adaptive control~\citep{aastrom2013adaptive} aims to improve control performance by adapting model parameters continuously. Under certain persistent excitation conditions and for certain system classes, convergence to a reference trajectory and even parameter convergence can be guaranteed. However, divergence is a serious concern during nominal operation, where the persistent excitation conditions are not necessarily satisfied.  This can often be a significant issue in practice. 
Here, we propose a different concept: rather than adapting or learning parameters all the time, we make a specific decision about \emph{when to learn} a new model or parameters. 
By separating learning from nominal behavior, we ensure resource efficiency and informative data, which is collected in dedicated learning experiments.

Adaptive filtering, change, fault, and anomaly detection have been developed to identify defective and malfunctioning systems for a variety of applications (see~\citep{gustafsson2000adaptive, gertler2013fault, isermann1984process, isermann2006fault} for an overview). In this article, we develop a method that is conceptually related to these research areas but closely tailored to NCSs with intermittent communication. In ETL, we combine tools from stochastic calculus and statistics to predict and analyze inter-communication times. 
Inter-communication times are independent, identically distributed, and scalar-valued. Due to these advantageous statistical properties, we can provide guarantees for the developed learning triggers.

Comparing expected behavior with observed realizations is also common in cognitive science to model human learning~\citep{butz2003anticipatory}.
This effect is often quantified with the aid of internal models and anticipation along surprisal boundaries~\citep{butz2003internal}. Interestingly, these ideas are very similar to ETL from an abstract point of view. However, the considered systems, concrete implementation, and developed theory differ significantly.
Notwithstanding, there exists a conceptual connection to learning in biological systems. 
\subsection*{Contributions}

We propose the novel idea of event-triggered learning and develop the concept in the context of networked control systems.  In particular, we derive data-driven learning triggers based on statistical properties of inter-communication times. These triggers make principled decisions on \emph{when} to learn a new model.

In  detail,  this  article  makes  the  following
contributions:
\begin{itemize}
    \item introducing event-triggered learning with probabilistic guarantees ensuring the effectiveness of the proposed learning triggers; 
    \item treatment of both cases, perfect state measurements and output measurements (Kalman filtering), and analysis of their differences; and
    \item demonstration of improved prediction accuracy and reduced communication in numerical simulations.
\end{itemize}
Preliminary results were presented in the conference paper~\citep{so18}.  This article significantly extents this paper and presents the full ETL framework.  While~\citep{so18} considered only learning triggers based on the expected value and the perfect state measurement case, we generalize to improved triggers based on the full distribution and also extent to the more relevant case of output measurements (Kalman filtering).  Furthermore, the theoretical properties of the sample-based triggers are improved, and new illustrative examples are presented.

\section{Problem Formulation}
\label{sec:probformulation}
In this section, we make the problem of event-triggered learning precise for linear Gaussian time-invariant systems. 
The framework is developed in the context of NCSs and primarily focuses on limited bandwidth.
Information exchange over networks is abstracted to be ideal in the sense that there are no packet drops or delays.
First, we state the problem formulation for continuous time systems. We then address the discrete time case separately since the technical details differ slightly. In \sect \ref{sec:filtering}, the problem is extended to output measurements and, in particular, the Kalman filter setting.
\subsection{Continuous Time Formulation}
Let $(S, \mathcal{F}, (\mathcal{F}_t)_{t \in \mathbb{R}^+}, \mathbb{P})$ be a filtered probability space and $X(t) \in \mathbb{R}^n$ a stochastic process, indexed by time $t\geq 0$.
Furthermore, assume $X(t)$ (\cf `Process' block in \fig \ref{fig:ETL-fullstate}) is a solution to the following linear stochastic differential equation (SDE)
\begin{equation}
 	\mathrm{d}X(t) = \mathcal{A} X(t) \mathrm{d}t+ \varc \mathrm{d}W(t),\quad X(0) = x_0.
	\label{eq:diffusionSDE}
\end{equation}
Solutions to the SDE \eqref{eq:diffusionSDE} are well investigated and also known as Ornstein-Uhlenbeck (OU) processes~\citep{oksendal2003stochastic}. Further, let $\mathcal{A} \in \mathbb{R}^{n\times n}$ be a matrix with negative eigenvalues, which may in practice be obtained by applying local feedback control and considering the stable closed-loop dynamics. Assume $\varc \in \mathbb{R}^{n\times n}$ is a positive definite matrix, $W(t) \in \mathbb{R}^n$ a standard Wiener process that models process noise, and the initial point $x_0 \in \mathbb{R}^n$ is known. We denote the system parameters as $\theta = (\mathcal{A}, \varc)$ and models as $\hat{\theta} = (\hat{\mathcal{A}}, \hat{\varc})$.

For the model-based predictions (`Model-based Predictions' in \fig \ref{fig:ETL-fullstate}), we use the expected value of system \eqref{eq:diffusionSDE}, which coincides with the open-loop predictions of the deterministic system $\mathrm{d}\check{X}(t) = \hat{\mathcal{A}} \check{X}(t) \mathrm{d}t$, with $\check{X}(0) = x_0$.
Due to the stochasticity of the system, the prediction error will almost surely leave any predefined domain after sufficient time.
Event-triggered communication (`State Trigger' in \fig \ref{fig:ETL-fullstate}) bounds the prediction error by resetting the open-loop predictions $\check{X}(t)$ to the current state $X(t)$. Further, the binary event trigger
\begin{equation}
    \gamma_{\mathrm{state}}=1 \iff \| X(t) - \check{X}(t) \|_2 \geq \delta,
    \label{trigger:ETSE}
\end{equation}
is only activated when the error threshold $\delta>0$ is crossed and hence, limits communication to necessary instances.
The corresponding inter-communication time is defined as
\begin{equation}
\stcts \coloneqq \inf \{t \in \mathbb{R} : \| X(t) - \check{X}(t) \|_2 \geq \delta\},
\label{eq:ST}
\end{equation}
and realizations of this random variable are denoted as~$\stcts_1,\ldots,\stcts_n$ and can be directly measured as the time between communication instances. 
\begin{assum}
\label{assum:bounded}
We assume $\tau \leq \tau_\mathrm{max} < \infty$.
\end{assum}
Bounded communication times are usually implemented in real-world applications to detect defect agents, which never communicate.
Hence, communication is enforced after $\tau_\mathrm{max}$.
For the design of the final learning trigger to be derived in this article, the assumption can be omitted. However, it is useful for intermediate results, such as the expectation-based learning trigger.

We address the problem of designing learning triggers (`Learning Trigger' in \fig \ref{fig:ETL-fullstate}) based on inter-communication time analysis. Since the probability distribution of $\tau$ can be fully parameterized by $\theta$, we can derive an expected distribution based on the model $\hat{\theta}$ and test if empirical inter-communication times are drawn from that distribution.
Further, this statistical analysis yields theoretical guarantees, which are obtained from concentration inequalities and ensure that the derived learning triggers are effective. Therefore, we design a method to perform dedicated learning experiments on necessity and update models $\hat{\theta}$ in an event-triggered fashion.

\subsection{Discrete Time Formulation}
\label{sec:disctime}
Since processing on microcontrollers or sensors mostly happens on synchronously sampled data, we provide an alternative discrete time formulation of the considered problem.
In principle, the problem formulation does not change.
However, some essential details differ; for example, the inter-communication times from \eqref{eq:ST} need to be treated differently due to discontinuities in the states.

The discrete time analogue to \eqref{eq:diffusionSDE} is
\begin{equation}
        x(k+1) = Ax(k)+ \epsilon(k),\quad x(0) = x_0,
        \label{eq:sysdisc}
\end{equation}
with discrete time index $k\in\mathbb{N}$ and state $x(k) \in \mathbb{R}^n$. Furthermore, we assume $A \in \mathbb{R}^{n\times n}$ has all eigenvalues strictly within the unit sphere and $\epsilon(k) \sim \mathcal{N}(0 , \vard)$ with $\vard \in \mathbb{R}^{n\times n}$ being symmetric and positive definite. 
The model-based predictions are obtained through the equation~$\check{x}(k+1) = \hat{A} \check{x}(k)$, which yields the trigger
\begin{equation}
    \gamma_{\mathrm{state}}=1 \iff \| x(k) - \check{x}(k) \|_2 \geq \delta.
    \label{trigger:ETSEdisc}
\end{equation}
We define the system parameters and model as $\theta = (A, \vard)$ and $\Hat{\theta} = (\hat{A},\Hat{\vard})$.  
Hence, we obtain the inter-communication times
\begin{equation}
\stdisc \coloneqq \min \{k \in \mathbb{N} : \| x(k) - \check{x}(k) \|_2 \geq \delta\}.
\label{eq:emp-ST-disc}
\end{equation}

\section{Communication as Stopping Times}
\label{Sec:ST}
In this section, we characterize inter-communication times (Eq. \eqref{eq:ST}) as stopping times of the prediction error process. The inter-communication time $\tau$ is a random variable and depends on the stochastic system \eqref{eq:diffusionSDE}. 
We seek to compare model-based expectations to observed data in order to detect significant inconsistencies between $\theta$ and $\hat{\theta}$.
The core idea of the learning triggers comes down to deriving expected stopping time distributions based on the model $\hat{\theta}$ and then analyzing how likely it is that observed stopping times $\tau_1,\ldots,\tau_n$ are drawn from this distribution.

Assuming $\hat{\theta} = \theta$, we derive model-based statistical properties of $\tau$. 
Later on, we will test the hypothesis that empirical inter-communication times are indeed drawn from the derived distribution of $\tau$---if not, this will indicate  $\hat{\theta} \not= \theta$; that is, the model does not match reality.

\subsection{Theoretical Properties}

We define the error process as $Z(t) \coloneqq X(t) - \check{X}(t)$.
Due to linearity, it follows immediately that $Z(t)$ is an OU process as well and that $Z(0) = 0$.
Next, we introduce inter-communication times with respect to the stochastic process $Z(t)$.
Assume $\mathcal{F}_t = \sigma(Z_s : s\leq t)$ is the natural filtration on the given probability space and $\tau$ a stopping time with respect to $\mathcal{F}_t$.
In particular, we consider the first exit time of the stochastic process $Z(t)$ from a sphere with radius $\delta$, \ie $\tau \coloneqq \inf \{t \in \mathbb{R} : \|Z(t)\|_2 > \delta\}$, 
which precisely coincides with \eqref{eq:ST}. Hence, we use the terms stopping times and inter-communication times synonymously in this article.

After each communication instance, we reset the process $Z(t)$ and set it to zero again by correcting $\check{X}(t)$ to $X(t)$.
The sample paths of the process $Z(t)$ are (almost surely) continuous between two inter-communication times, which follows from the existence and uniqueness theorem of solutions to SDEs (\cf~\citep{oksendal2003stochastic}). Therefore, we can precisely quantify the moment when the error threshold \eqref{trigger:ETSE} is crossed. 
Further, it is possible to quantify statistical properties of $\tau$ such as the expected value~\citep[\sect 7.2]{Pavliotis:2014} or the distribution~\citep{patie2008first} with the aid of certain boundary value problems. In particular, there are existence and uniqueness theorems~\citep{patie2008first} that imply that $\tau$ is mathematically well behaved.

\subsection{Monte Carlo Approximations}
Next, we describe how we obtain statistical properties of $\tau$ such as expected value $\mathbb{E}[\tau]$, variance $\mathbb{V}[\tau]$, and cumulative distribution function (CDF) $F(t)$ with the aid of sample-based methods and hence, without solving nonlinear boundary value problems. 
Given the system parameters $\theta$, we can simulate trajectories of the stochastic process \eqref{eq:diffusionSDE} with the aid of numerical sample methods such as the Euler-Maruyama scheme (\cf~\citep[\sect 10.2]{kloeden2011numerical} for an introduction to numerical solutions of SDEs).

In order to obtain independent and identically distributed (i.i.d.) samples $\tau_1,\ldots,\tau_n$, we sample the process $Z(t)$ and restart from zero after reaching the threshold $\delta$. Alternatively, we could also simulate $X(t)$ and $\check{X}(t)$ and set the predictions $\check{X}(\tau)$ to the true value $X(\tau)$ when communication is triggered. The statistical properties of the corresponding stopping times do not differ because the processes $Z(t)$ and $X(t) - \check{X}(t)$ are indistinguishable. Further, stable OU processes are stationary and satisfy the strong Markov property, which generalizes the Markov property to stopping times.

For given i.i.d.\ random variables, we can approximate the expected value with $\frac{1}{n}\sum_{i=1}^n \tau_i$ and the CDF with $F_n(t) \coloneqq\frac{1}{n}\sum_{i=1}^n \mathbbm{1}_{\tau_i \leq t}$, where $\mathbbm{1}$ is the indicator function. 
Quantifying the convergence speed of the above approximation will be vital in designing learning triggers.

\section{Learning Trigger Design for Continuous Time}
In this section, we design the learning trigger $\gamma_\mathrm{learn}$ (\cf \fig \ref{fig:ETL-fullstate}) to detect a mismatch between model and true dynamics based on the inter-communication time~$\tau$.

\subsection{Concentration Inequalities}
The following results will form the backbone of the later derived learning triggers. Concentration inequalities quantify the convergence speed of empirical distributions to their analytical counterparts. In particular, Hoeffding's inequality bounds the expected deviation between mean and expected value. Further, we also consider the Dvoretzky-Kiefer-Wolfowitz (DKW) inequality, which compares empirical and analytical CDF functions, and bounds the error between them uniformly.
Essentially, we test if observed data fits the distribution, which is induced by the model $\Hat{\theta}$; that is, which was derived with $\hat{\theta} = \theta$ (\cf \sect \ref{Sec:ST}).
If the distributions do not match, we conclude an unfit model and update $\Hat{\theta}$ through model learning.
\begin{lem}[\citet{hoeffding1963probability}]
 Let $\tau_1,\ldots, \tau_n$ be i.i.d.\ bounded random variables, s.\,t. $\tau_i \in [0,\tau_\mathrm{max}]$. Then
 \begin{equation}
     \mathbb{P}\left[\left| \frac{1}{n}\sum^n_{i=1} \tau_i - \mathbb{E}[\tau]\right|  > \kappa \right] \leq 2 \exp\left(-\frac{2n\kappa^2}{\tau^2_\mathrm{max}}\right).
    \label{eq:hoeff}
 \end{equation}
\end{lem}
We will first design learning triggers around the Hoeffding's inequality and later move on to richer statistical information.
Therefore, we also want to analyze the convergence speed of the empirical CDF function.
\begin{lem}[DKW Inequality~\citep{massart1990tight}]
\label{thm:DKW}
Assume $\tau_1,\ldots,\tau_n$ are i.i.d.\ random variables with CDF $F(t)$ and empirical CDF $F_n(t)$. Then
\begin{equation}
\mathbb{P}\left[\sup\limits_{t \in \mathbb{R}} |F_n(t) - F(t) | > \kappa \right] \leq 2\exp(-2n\kappa^2).
\label{eq:DKW}
\end{equation}
\end{lem}
\subsection{Expectation-based Learning Trigger}
We propose a first learning trigger $\gamma_\mathrm{learn}$ based on the expected value $\mathbb{E}[\tau]$.

\subsubsection{Exact Learning Trigger}
Based on the foregoing discussion, we propose the following learning trigger:
\begin{equation}
\gamma_{\mathrm{learn}} = 1 \iff	\left| \frac{1}{n}\sum\limits_{i=1}^n \tau_i - \mathbb{E}[\tau] \right| \geq \kappa_{\mathrm{exact}},
	\label{trigger}
\end{equation}
where $\gamma_{\mathrm{learn}} = 1$ indicates that a new model shall be learned; $\mathbb{E}[\tau]$ is the analytical expected value, which is based on the model $\Hat{\theta}$; and $\tau_1, \tau_2, \dots, \tau_n$ are the last $n$ empirically observed inter-communication times ($\tau_i$ the duration between two state triggers \eqref{trigger:ETSE}).
The horizon $n$ is chosen to yield robust triggers in the sense that a larger time horizon allows the detection of smaller changes. However, it also increases the delay until the $n$ samples are actually observed. 
The threshold parameter $\kappa_{\mathrm{exact}}$ quantifies the error we are willing to tolerate. 
There are some examples where it is possible to compute $\mathbb{E}[\tau]$ analytically. In general, however, it is intractable. 
Hence, we also propose the approximated learning trigger, which takes the approximations for the statistical analysis into account.
We denote \eqref{trigger} as the \emph{exact learning trigger} because it involves the exact expected value $\mathbb{E}[\tau]$, as opposed to the trigger derived in the next subsection, which is based on a Monte Carlo approximation of the expected value.

Even though the trigger \eqref{trigger} is meant to detect inaccurate models, there is always a chance that the trigger fires not due to an inaccurate model, but instead due to the randomness of the process (and thus randomness of inter-communication times $\tau_i$).  
Such false positives are inevitable due to the stochastic nature of the problem.  
However, we obtain a confidence interval, which contains the empirical mean with high confidence. If observations violate the derived confidence interval, we conclude that distributions do not match, and learning is beneficial.
Therefore, we propose to choose $\kappa_{\mathrm{exact}}$ to yield effective triggering with a user-defined confidence level.
We then have the following result for the trigger \eqref{trigger}:
\begin{thm}[Exact learning trigger] 
\label{thm:exactTrigger}
Assume 
\newline
$\tau$ and $\tau_1,\ldots,\tau_n$ are i.i.d.\ random variables and the parameters $\alpha$, $n$, and $\tau_{\mathrm{max}}$ are given.
If the trigger \eqref{trigger} gets activated ($\gamma_{\mathrm{learn}} = 1$) with
\begin{equation}
\kappa_{\mathrm{exact}} = \tau_{\mathrm{max}}\sqrt{\frac{1}{2n} \ln \frac{2}{\alpha} },
\label{eq:kappa_exact}
\end{equation}
then
\begin{equation}
	\mathbb{P}\left[\left| \frac{1}{n}\sum\limits_{i=1}^n \tau_i -\mathbb{E}[\tau] \right|\geq \kappa_{\mathrm{exact}} \right] \leq \alpha.
\end{equation}
\end{thm}
\begin{pf}
	Substituting $\kappa_{\mathrm{exact}}$ into the right-hand side of Hoeffding's inequality yields the desired result. \myqed
\end{pf}
The theorem quantifies the expected convergence rate of the empirical mean to the expected value for a perfect model.
This result can be used as follows: the user specifies the desired confidence level $\alpha$, the number $n$ of inter-communication times considered in the empirical average, and the maximum inter-communication time $\tau_{\mathrm{max}}$.  By choosing $\kappa_{\mathrm{exact}}$ as discussed, the exact learning trigger \eqref{trigger} is guaranteed to make incorrect triggering decisions (false positives) with a probability of less than $\alpha$.  
\subsubsection{Approximated Learning Trigger}
As discussed in \sect \ref{Sec:ST}, obtaining $\mathbb{E}[\tau]$ can be difficult and computationally expensive. Instead, we propose to approximate $\mathbb{E}[\tau]$ by sampling $\tau$. For this, we simulate the process $Z(t)$ and average the obtained stopping times $\hat{\tau}_1,\ldots, \Hat{\tau}_m$. 
This yields the \emph{approximated learning trigger}
\begin{equation}
\gamma_{\mathrm{learn}} = 1 \iff	\left| \frac{1}{n}\sum\limits_{i=1}^n \tau_i - \frac{1}{m}\sum\limits_{i=1}^m \Hat{\tau}_i \right| \geq \kappa_{\mathrm{approx}}.
\label{approxtrigger}
\end{equation}

The Monte Carlo approximation leads to a choice of $\kappa_{\mathrm{approx}}$, which is different from $\kappa_{\mathrm{exact}}$ for small $m$. For $m\to \infty$ we see that $\kappa_{\mathrm{approx}}$ converges to $\kappa_{\mathrm{exact}}$.
\begin{thm}[Approximated Learning Trigger]
\label{Thrm:MLT}
Assume $\tau_1,\ldots,\tau_n,$ and $\hat{\tau}_1,\ldots,\hat{\tau}_m$ are i.i.d.\ random variables.
If the trigger \eqref{approxtrigger} gets activated ($\gamma_{\mathrm{learn}} = 1$) with 
\begin{equation}
\kappa_{\mathrm{approx}} = \tau_\mathrm{max}\sqrt{\frac{n+m}{2nm}\ln{\frac{2}{\alpha}}},
\label{kappa:Happrox}
\end{equation}
then
\begin{equation}
\mathbb{P}\left[\left| \frac{1}{n}\sum\limits_{i=1}^n \tau_i - \frac{1}{m}\sum\limits_{i=1}^m \Hat{\tau}_i \right| \geq \kappa_{\mathrm{approx}}\right] \leq \alpha.
\end{equation}
\end{thm}
\begin{pf}
First, we introduce an alternative formulation of Hoeffding's inequality \eqref{eq:hoeff}
\begin{align*}
    \mathbb{P}\left[\left| \frac{1}{n}\sum^n_{i=1} \tau_i - \frac{1}{m}\sum^m_{i=1} \hat{\tau}_i - \left( \mathbb{E}[\tau] - \mathbb{E}[\hat{\tau}] \right) \right| > \kappa_{\mathrm{approx}}\right] \\
    \leq 2\exp\left(-\frac{2\kappa_{\mathrm{approx}}^2}{(m^{-1}+n^{-1}) \tau_\mathrm{max}^2}\right),
\end{align*}
which was already stated in the original article by Hoeffding~\citep{hoeffding1963probability} as a corollary (Eq. 2.6). Here, we assume that $\tau$ and $\hat{\tau}$ are identically distributed and, therefore, the analytical expected values cancel out. Rearranging for $\kappa_{\mathrm{approx}}$ yields the desired result. \myqed
\end{pf}
\subsection{Density-based Learning Trigger}
Analyzing the expected values is, in general, not enough to distinguish random variables since higher moments such as variance can differ. Therefore, we propose to look at the CDF, build learning triggers around the DKW inequality \eqref{eq:DKW}, and thus use richer statistical information. We propose the following learning trigger:
\begin{equation}
     \gamma_{\mathrm{learn}}=1 \iff \sup\limits_{t \in \mathbb{R}} |F(t) - F_n(t) | > \kappa_{\mathrm{exact}}.
     \label{trigger:KSonesided}
\end{equation}
The density-based learning trigger has the following property:
\begin{thm}[Exact Density Learning Trigger] 
\label{thm:exactTriggerKS}
Assume  $\tau_1,\ldots,\tau_n$ are i.i.d.\ random variables with CDF $F(t)$ and empirical CDF $F_n(t)$. If the learning trigger \eqref{trigger:KSonesided} gets activated ($\gamma_{\mathrm{learn}} = 1$) with
\begin{equation}
    \kappa_{\mathrm{exact}} =\sqrt{\frac{1}{2n} \ln \frac{2}{\alpha} },
\end{equation}
then 
\begin{equation}
    \mathbb{P}\left[\sup\limits_{t \in \mathbb{R}} |F(t) - F_n(t) | > \kappa_{\mathrm{exact}}\right] \leq \alpha.
\end{equation}
\end{thm}
\begin{pf}
Follows directly from the DKW Inequality.
\myqed
\end{pf}
Finally, we can follow the reasoning as before and obtain the sample-based version of the trigger \eqref{trigger:KSonesided}
\begin{equation}
     \gamma_{\mathrm{learn}}=1 \iff \sup\limits_{t \in \mathbb{R}} |\Hat{F}_m(t) - F_n(t) | > \kappa_{\mathrm{approx}},
     \label{trigger:KStwosided}
\end{equation}
where $\Hat{F}_m(t)= \frac{1}{m}\sum_{i=1}^m \mathbbm{1}_{\Hat{\tau}_i \leq t}$ and $\Hat{\tau}_i$ are obtained by creating samples based on the model $\Hat{\theta}$.
This trigger is essentially the well established two-sample Kolmogorov-Smirnov (KS) test~\citep{hodges1958significance}.
\begin{thm}[Two-sample KS Learning Trigger]
\label{thm:KStest}
Assume  $\tau_1,\ldots,\tau_n$and $\hat{\tau}_1,\ldots,\hat{\tau}_m$ are i.i.d.\ random variables with empirical CDFs $F_n(t)$ and $\Hat{F}_m(t)$. If the trigger
\eqref{trigger:KStwosided} gets activated with
\begin{equation}
    \kappa_{\mathrm{approx}} = \sqrt{\frac{n+m}{2nm}\ln\left(\frac{2}{\alpha}\right)},
\label{kappa:KSapprox}
\end{equation}
then 
\begin{equation}
    \mathbb{P}\left[\sup\limits_{t \in \mathbb{R}} |\Hat{F}_m(t) - F_n(t) | > \kappa_{\mathrm{approx}}\right] \leq \alpha.
\end{equation}
\end{thm}
\begin{pf}
 Follows from the two-sample KS test.
 \myqed
\end{pf}
The density-based learning triggers do not depend on $\tau_\mathrm{max}$ and consider richer statistical information, which can be an advantage and will be discussed in detail in the experimental sections. 
\section{Learning Trigger Design for Discrete Time}
Based on the previous discussion, we will now highlight how to apply the derived learning triggers to discrete time systems \eqref{eq:sysdisc}.
The random variables $\stcts$ (\cf \eq\eqref{eq:ST}) and $\stdisc$ (\cf \eq \eqref{eq:emp-ST-disc}) can differ significantly due to discretization effects. Intuitively, this effect can be thought of as the continuous time process crossing the $\delta$-threshold and returning within the discretization time. Therefore, the discrete-time process has no possibility of observing the crossing, and hence, stopping times tend to be larger for discrete time systems. For small time steps, the difference tends to be negligible, and $\stdisc$ converges to $\stcts$ in the limit. 

In this section, we show that the \emph{approximated} learning triggers transfer without any modification to the discrete time system. It is important to adjust the system parameters $\theta = (A, \vard)$ and the model $\hat{\theta} = (\hat{A}, \hat{\vard})$  to discrete time (\cf \sect \ref{sec:disctime}) in order to sample from the correct distribution (\ie sampling from the continuous time model, while the true dynamics are discrete, or vice versa). Only based on statistical tests, irrelevant of the actual shape, we decide if they coincide.  
\begin{thm}[Discrete Time Learning Trigger]
Assume $\theta$ and $\hat{\theta}$ correspond to the discrete time system \eqref{eq:sysdisc}. Then, the previously derived approximated learning triggers \eqref{approxtrigger} and \eqref{trigger:KStwosided} are applicable without any further modification.
\end{thm}
\begin{pf}
The derived learning triggers test if given observations of inter-communication times $\tau_1, \ldots, \tau_n$ are drawn from the same distribution as $\hat{\tau}_1,\ldots,\hat{\tau}_m$ and therefore, if $\theta = \hat{\theta}$. The concrete shape of the distribution is irrelevant for the test.
\myqed
\end{pf}
\begin{rem}
It is also possible to consider more complex noise models such as colored noise. 
The main challenge lies in identifying the system and, in particular, the noise model from data. 
\end{rem}

\section{Numerical Example -- Reduced Communication} 
\label{sec:numExample}
The learning triggers derived in the previous two sections are the core element in the proposed ETL architecture (\fig \ref{fig:ETL-fullstate}, block `Learning Trigger').  For `Model Learning' in the context of linear Gaussian systems considered herein, one can use standard techniques for linear systems identification~\citep{ljung1999system}, which we do not elaborate further.  Thus, all components of the proposed ETL method in \fig \ref{fig:ETL-fullstate} are complete, and we present a first numerical example to illustrate the main ideas of the developed learning triggers.
{\scriptsize
\begin{figure}[tb]
	\begin{center}
		\includegraphics[width=0.45\textwidth]{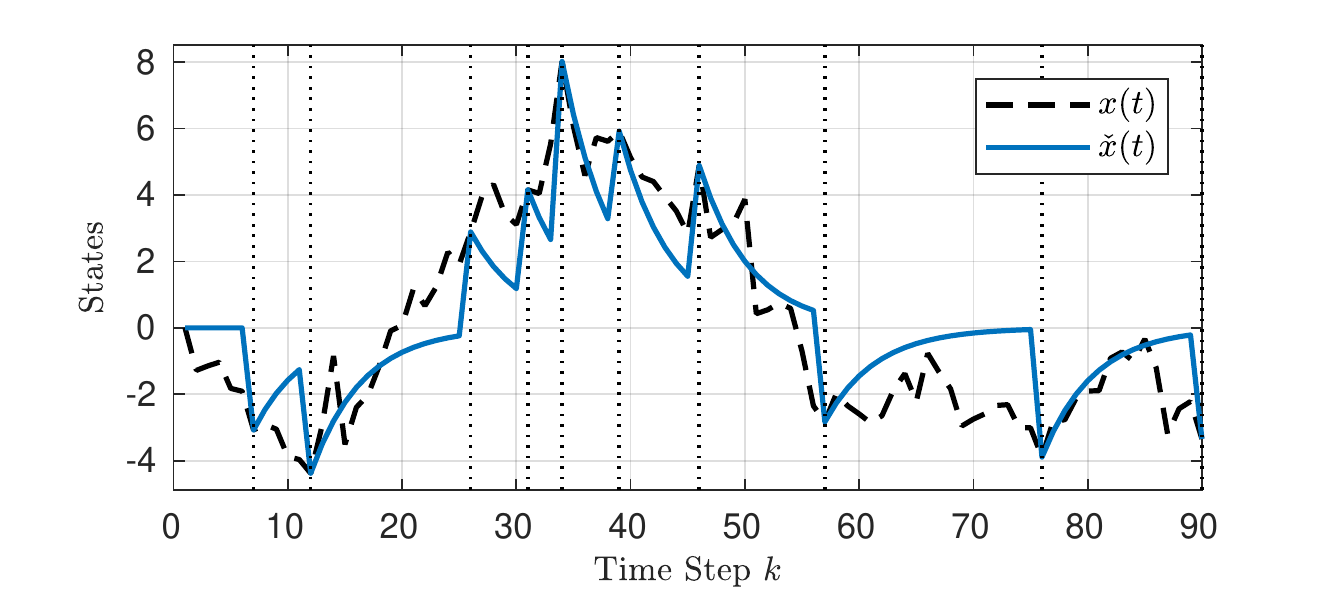}\\
		\includegraphics[width=0.45\textwidth]{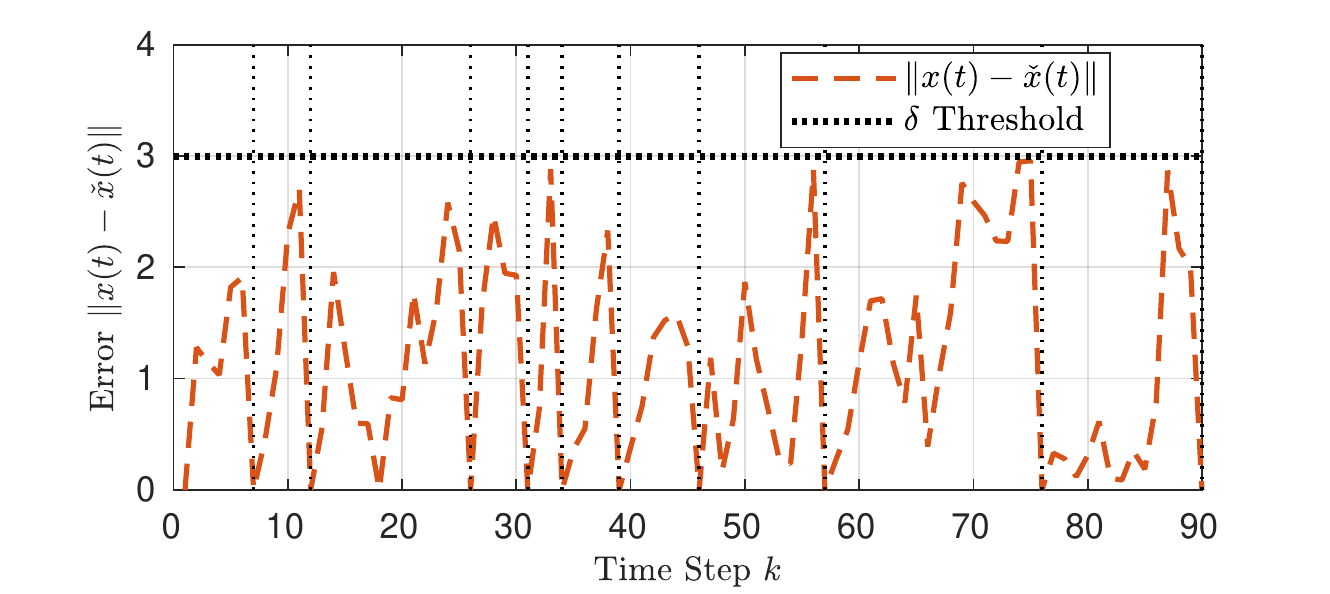}\\
		\includegraphics[width=0.45\textwidth]{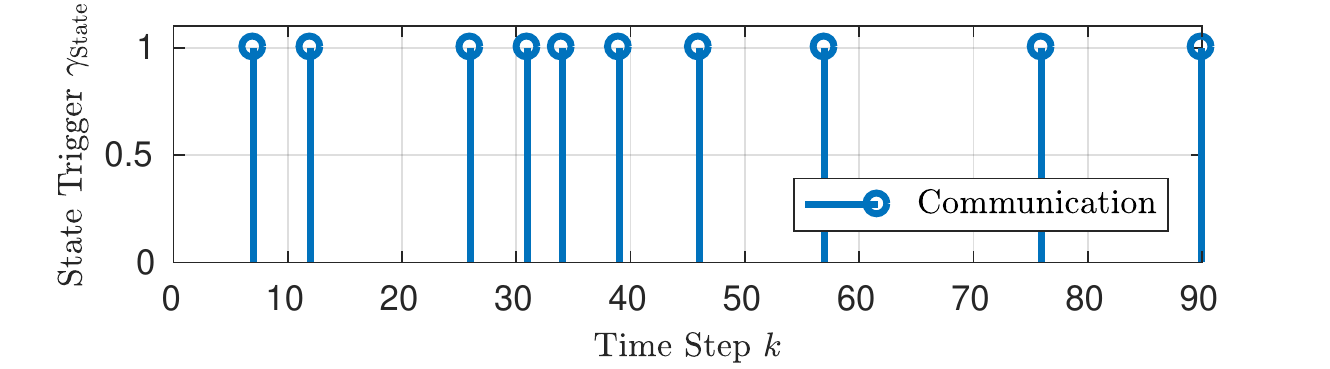}
		\caption{\scriptsize Example of model-based state predictions, which are reset to the exact state whenever the error would exceed the predefined threshold $\delta =3$. States, error signals, and the first ten communication instances $\tau_i$ are depicted in the three graphs (top to bottom).}
		\label{fig:states}
	\end{center}
\end{figure}}

\begin{figure}[tb]
	\begin{center}
		\includegraphics[width=0.45\textwidth]{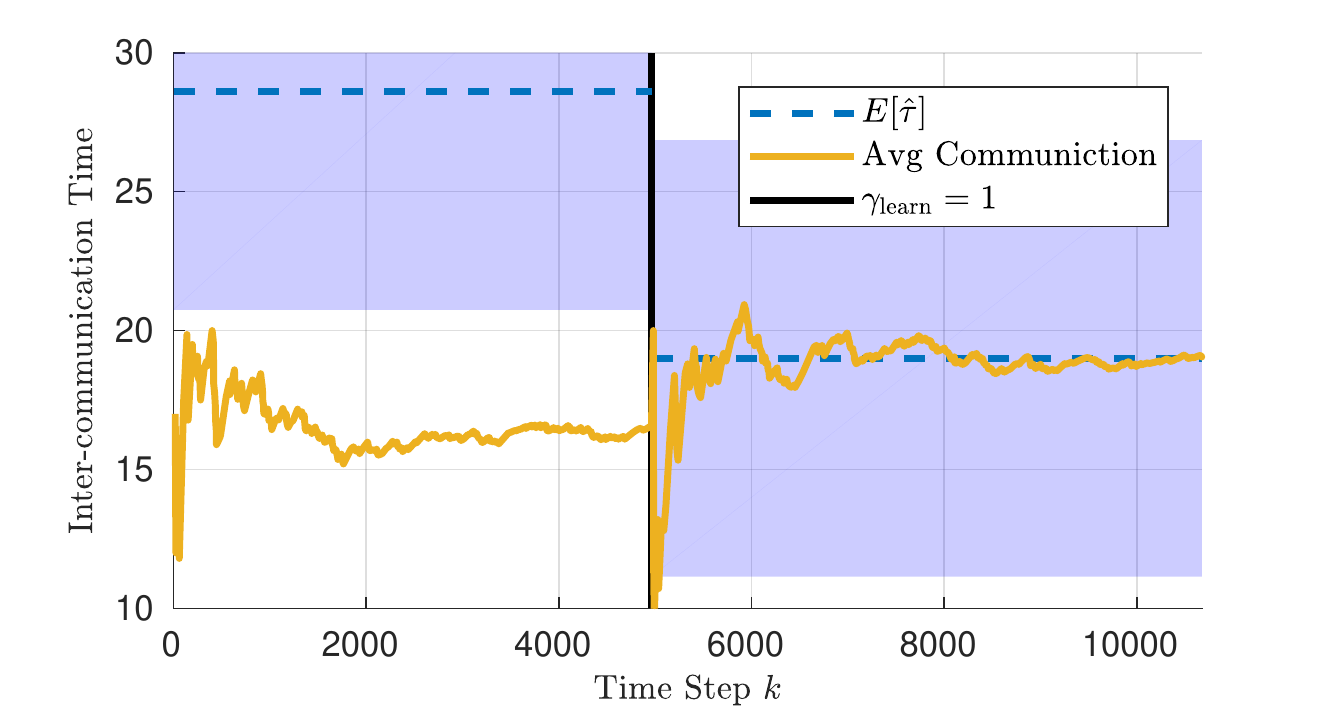}
		\caption{\scriptsize Average inter-communication times. The dashed blue line illustrates the model-based expected value and the yellow line the empirical mean. The shaded blue area around the dashed blue line indicates the confidence interval that should contain the empirical mean with 95\% probability. Every new communication instant is added to a buffer of size $n=300$. Afterward, the learning trigger compares the expected value with empirical mean and updates the model if the yellow line is outside the blue area. A new model-based expected value is computed, and the empirical mean coincides with it. Further, the inter-communication times increase, which corresponds to a decrease in communication.}
		\label{fig:avgcomm}
	\end{center}
\end{figure}

\subsection{Setup}
Next, we introduce the system, and afterward, we apply the learning trigger in order to demonstrate how to detect an inaccurate model. 
We consider the first-order dynamical system $x(k+1) = 0.9 x(k) + \epsilon(k)$, with noise $\epsilon(k) \sim \mathcal{N}(0, 1)$. 
Further, we assume the disturbed model $\hat{\theta} = (0.8, 1)$ and hence, we obtain the predictions with the equation $\check{x}(k+1) = 0.8 \check{x}(k)$.
To bound the prediction error, we deploy the state trigger \eqref{trigger:ETSEdisc} with $\delta = 3$.
In \fig \ref{fig:states}, we can see in the first graph a trajectory of states $x(k)$ as a black dashed line and the model-based predictions $\check{x}(k)$ in blue. Whenever $\gamma_{\mathrm{state}}=1$, we set $\check{x}(\tau)$ to $x(\tau)$. The error signal never crosses the $\delta$-threshold and is depicted in the second graph. 
The communication instances are shown in the third graph. The distances between two consecutive communication instances corresponds to the inter-communication times. Further, we set $\alpha = 0.05$, $\tau_{\rm{max}}= 100$, $n= 300$, and $m= \num{100000}$.

\subsection{Expectation-based Learning Trigger}
The proposed learning triggers analyze the statistical properties of the observed inter-communication times and compare them to model-induced quantities. For instance, in \fig \ref{fig:avgcomm}, the corresponding average inter-communication rate is shown (yellow line). The first $n$ inter-communication times are stored in a buffer, and the empirical mean is computed.
Based on the model $\hat{\theta}$ and with the aid of $m$ Monte Carlo simulations, we derive $\mathbb{E}[\hat{\tau}]\approx 28.6$ (dashed blue line). The model-based confidence interval is obtained with the aid of Theorem~\ref{Thrm:MLT}. After the buffer is successfully filled (at $k = 4961$), the learning trigger \eqref{approxtrigger} compares if the buffered average inter-communication rate (yellow line) lies outside the expected confidence interval, which is the case here.
Therefore, the learning trigger discovers an inaccurate model and triggers a learning experiment. Here, we abstract learning and set model $\hat{\theta}$ to the true parameters $\theta$. A more detailed discussion on the learning aspect of ETL can be found in~\citep{so18}, where we demonstrated the effectiveness of ETL in hardware experiments on a cart-pole system.

After updating the model (at $k = 4961$), we empty the buffer, start collecting new stopping times, and reset the average inter-communication time accordingly. 
This causes the initial fluctuation of the signal. However, we see fast convergence to the model-based expectation.
Further, the average inter-communication time was increased after updating the model (yellow line converges to a larger value), which results in less communication.

The test statistic is also depicted in \fig \ref{fig:learningtrigger} for the initial inaccurate model and in \fig \ref{fig:learningtriggertreusys} for the exact model. The dashed blue line again represents the model-based expected value, while the dashed red line depicts the empirical mean at the moment of triggering.

\subsection{Density-based Learning Triggers}
\begin{figure}[tb]
	\begin{center}
		\includegraphics[width=0.4\textwidth]{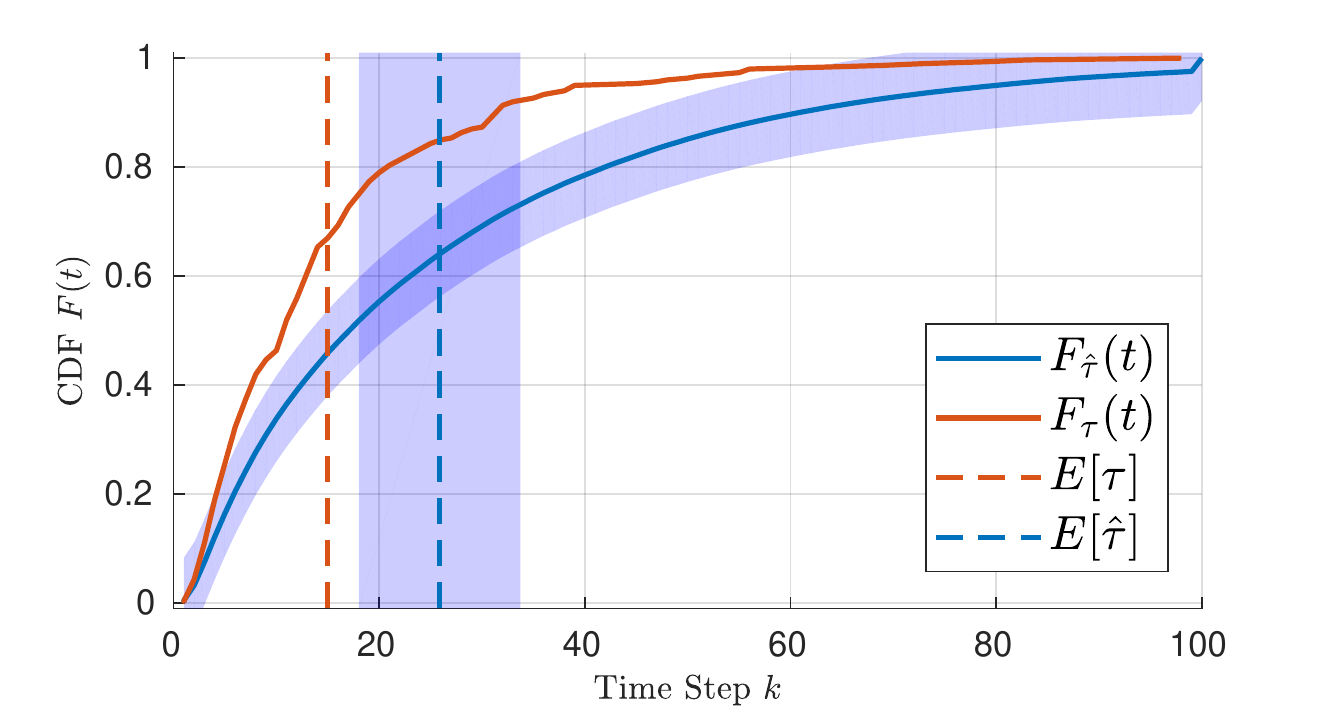}
		\caption{\scriptsize Statistical properties of expected communication $\Hat{\tau}$ (blue) and observed inter-communication times $\tau$ (red). The dashed lines capture the expected values, and the shaded blue region is a confidence interval that should contain the dashed red line with a 95\% probability (\cf \eqref{approxtrigger}). 
		As there is a significant deviation between observation and expectation, the learning trigger initiates to relearn the model. The solid lines represent the CDF functions, and also here, the empirical distribution is not contained within the confidence bounds, which triggers learning (\cf \eqref{trigger:KStwosided}).}
		\label{fig:learningtrigger}
	\end{center}
\end{figure}

\begin{figure}[tb]
	\begin{center}
		\includegraphics[width=0.4\textwidth]{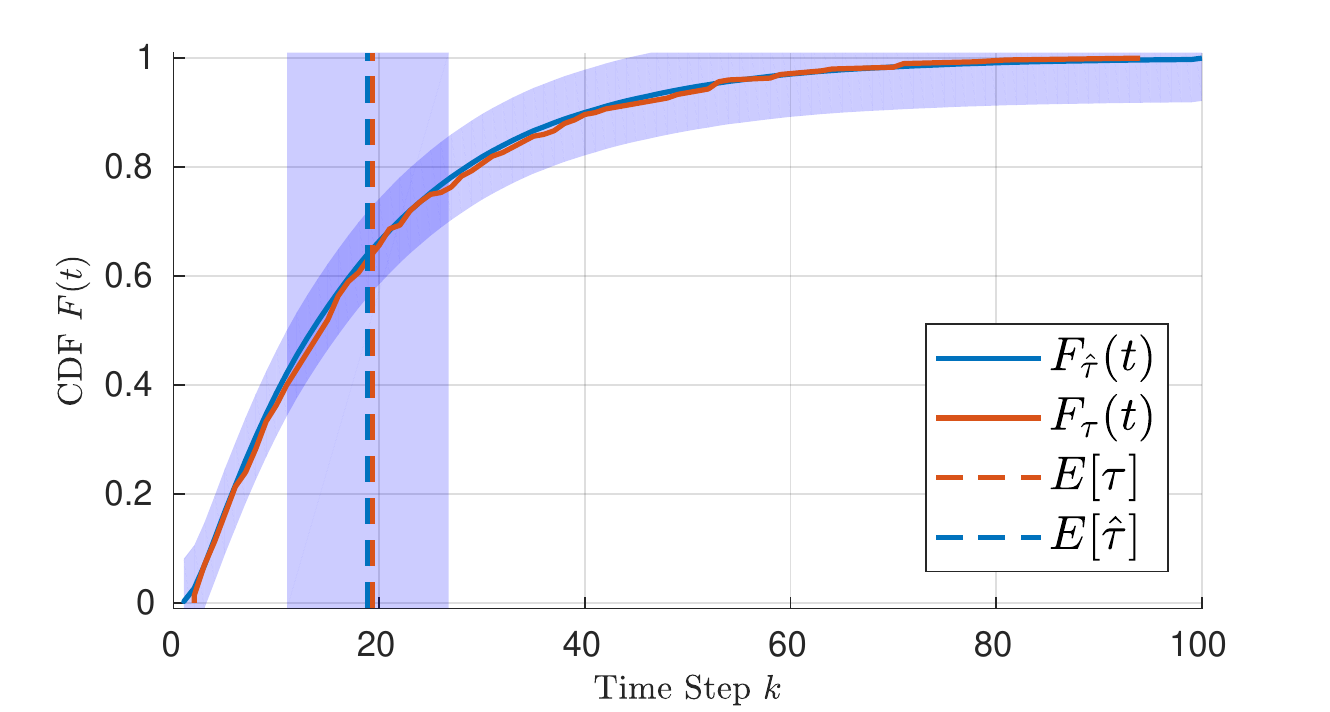}
		\caption{\scriptsize After updating the model $\hat{\theta}$ to the true system parameters $\theta$, the estimated stopping times $\hat{\tau}$ coincide with empirical stopping times $\tau$. As a direct consequence,  communication behavior is improved as the stopping times increase.}
		\label{fig:learningtriggertreusys}
	\end{center}
\end{figure}

Next, we will discuss the density-based learning trigger \eqref{trigger:KStwosided}, which is also illustrated in \fig \ref{fig:learningtrigger} and \fig \ref{fig:learningtriggertreusys}. The solid blue line represents the model-based CDF function $\hat{F}_m(t)$, and the solid red line is the empirical CDF $F_n(t)$ based on observed inter-communication times. Here, both triggers detect the inaccurate model (\cf \fig \ref{fig:learningtrigger}) and have high confidence in the true model since the model-based and empirical quantities coincide, which is depicted in \fig \ref{fig:learningtriggertreusys}.
The confidence interval is derived with Theorem \ref{thm:KStest} and tighter than the expectation-based. Hence, inaccurate models can be detected faster.
\subsection{Expected Value is not Enough}
\label{sec:momentmatch}
\begin{figure}[tb]
	\begin{center}
		\includegraphics[width=0.4\textwidth]{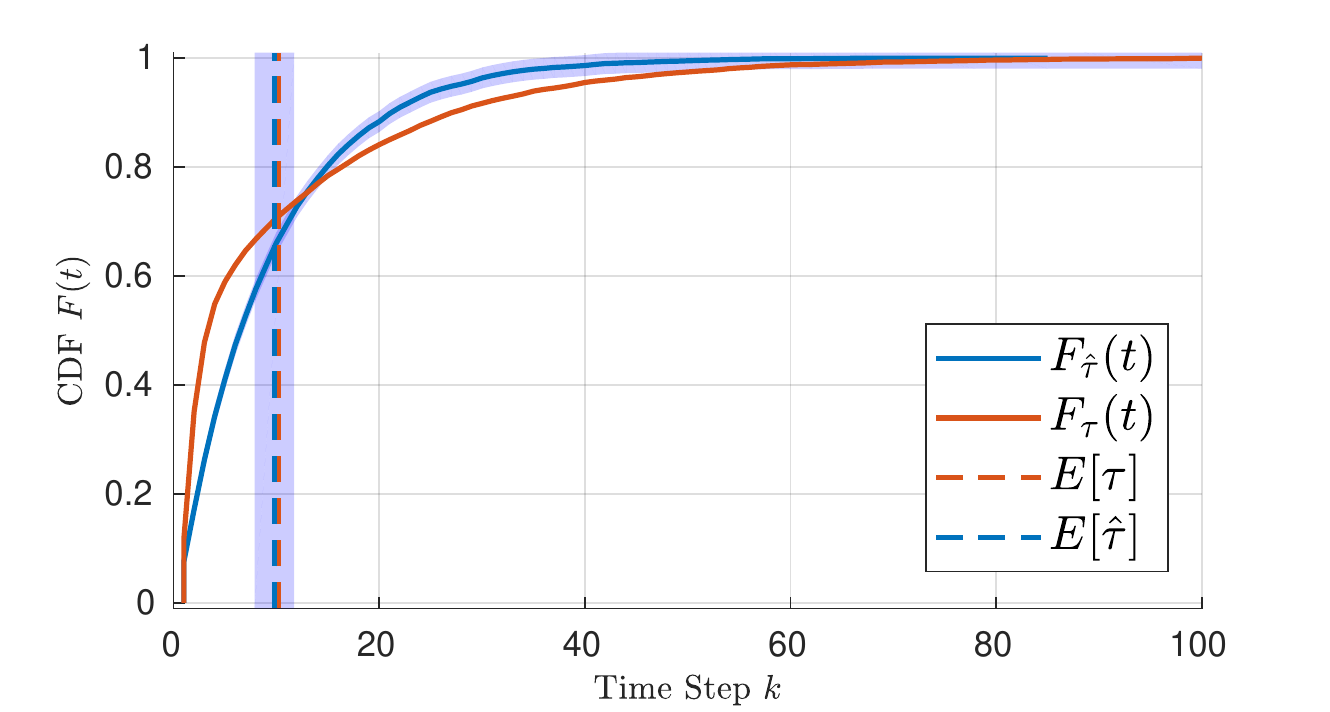}
	\end{center}
	\caption{Counterexample showing that the expected value of inter-communication times (learning trigger \eqref{approxtrigger}) may not be sufficient to detect an inaccurate model. Furthermore, the CDF-based trigger \eqref{trigger:KStwosided} detects the mismatch reliably.}
	\label{fig:countertest}
\end{figure}
Here, we assume the model $\Hat{\theta} = (0.5,1.7)$ with $n=m= \num{10000}$.
Intuitively, process noise 
and stability have opposite effects on the communication behavior.
We construct the counterexamples by creating a hypersurface of models $\Hat{\theta}$, where the noise and stability effects cancel out. 
Figure \ref{fig:countertest} shows the  expected and empirical expected values of inter-communication times, which are almost identical -- bad performance is expected and also realized due to the bad prediction model. The CDF-based trigger is still able to detect the inaccuracy, which is a big advantage over the expectation-based learning triggers. 
Clearly, the highest inter-communication time is realized when $\hat{\theta} = \theta$, which can be observed when comparing \fig \ref{fig:learningtriggertreusys} with \fig \ref{fig:countertest}, where the inter-communication time is twice as high.

Clearly, the model has also to be communicated at some point. However, this happens very rarely and, in particular, when there is a significant change in the system.
We conclude that both learning triggers are effective in detecting a mismatch between model and true dynamics. Also, average communication was successfully reduced after updating the model.

\section{Extensions and Insights}
So far, we assumed that perfect measures of the full state $x(k)$ are available at the sending agent. In the following, we will drop this assumption and consider systems where only part of the state can be measured.
\subsection{Output Measurements}
\label{sec:filtering}
Assume the following system
\begin{equation}
    x(k+1) = Ax(k)+ \epsilon(k),  \quad  y(k)= Cx(k) + \nu(k),
\label{eq:systrue}
\end{equation}
with output measurements $y(k)\in \mathbb{R}^m $.
Further, let $A \in \mathbb{R}^{n\times n}$ and $C \in \mathbb{R}^{n\times m}$. The system is again assumed to be stochastic with process noise $\epsilon(k) \sim \mathcal{N}(0 , \vard)$ and observation noise $\nu(k) \sim \mathcal{N}(0 , R)$, which are independent of each other.
We also assume that $A$ is stable, and the pair $(A,C)$ is observable.
Hence, the system is parameterized by $\theta = (A, C, \vard, R)$ and modeled by $\hat\theta = (\hat{A}, \hat{C} , \hat{\vard} , \hat{R})$.
To reconstruct the full state, we use a Kalman filter (KF), 
which is the optimal filter for linear Gaussian systems with exact models~\citep{anderson2012optimal}.
Here, we consider the steady-state KF and obtain
\begin{equation}
    \Hat{x}(k+1) = \hat{A} \Hat{x}(k) + K \left(y(k+1) - \hat{C}\hat{A}\Hat{x}(k)\right),
    \label{eq:sysKF}
\end{equation}
where $K \in \mathbb{R}^{m \times n}$ is the corresponding gain. Further, assume the process has already converged to stationarity. 

Ideally, we want to use the KF states $\hat{x}(k)$ on the receiving agent's site. However, this would require periodic communication of the estimates $\hat{x}(k)$, or the measurements $y(k)$, which we try to avoid. Exactly as in \sect \ref{sec:probformulation}, we run a model-based prediction step in the absence of data, obtain $\check{x}(k+1) = \hat{A} \check{x}(k)$, and employ the state trigger
$\| \hat{x}(k) - \check{x}(k) \|_2 \geq \delta$.
Hence, the inter-communication time is defined as $\stkf \coloneqq \min \{k \in \mathbb{N} : \| \hat{x}(k) - \check{x}(k) \|_2 \geq \delta\}$.
Extending ETL to output measurements is based on treating the KF sequence as a stochastic process in its own right and investigating the distribution of $\hat{x}(k)$.
With the aid of the innovation sequence, we can derive an auto-regressive structure.
The innovation of the KF is defined as $I(k) = y(k) - C \Hat{x}(k)$.
Furthermore, it is well known that $I(1),\ldots,I(n)$ are independent normal distributed random variables with $I(k) \sim \mathcal{N}(0 , S)$~\citep{anderson2012optimal}. The covariance is given by $S = CPC^\intercal + R$, where $P$ is the stationary error covariance matrix of the KF and can be obtained by solving the corresponding Riccati equation~\citep[Equation (4.4)]{anderson2012optimal}.
Hence, we reformulate the KF as
\begin{equation}
    \hat{x}(k+1) = A \hat{x}(k) + KI(k),
    \label{eq:KFinno}
\end{equation}
and regard $I(k) \sim \mathcal{N}(0 , S)$ as a random variable. 
By regarding $KI(k)$ as process noise, we are back to the previously discussed problem (\cf \eqref{eq:sysdisc}) and can apply the derived tools and learning triggers.
Hence, we can effectively analyze the distribution of the corresponding stopping time with the previously derived tools -- sampling \eqref{eq:KFinno} to obtain model-based stopping times $\Hat{\tau}^\mathrm{o}$.

\subsection{Better Models may Result in more Communication} 
\begin{figure}[tb]
	\begin{center}
		\includegraphics[width=0.4\textwidth]{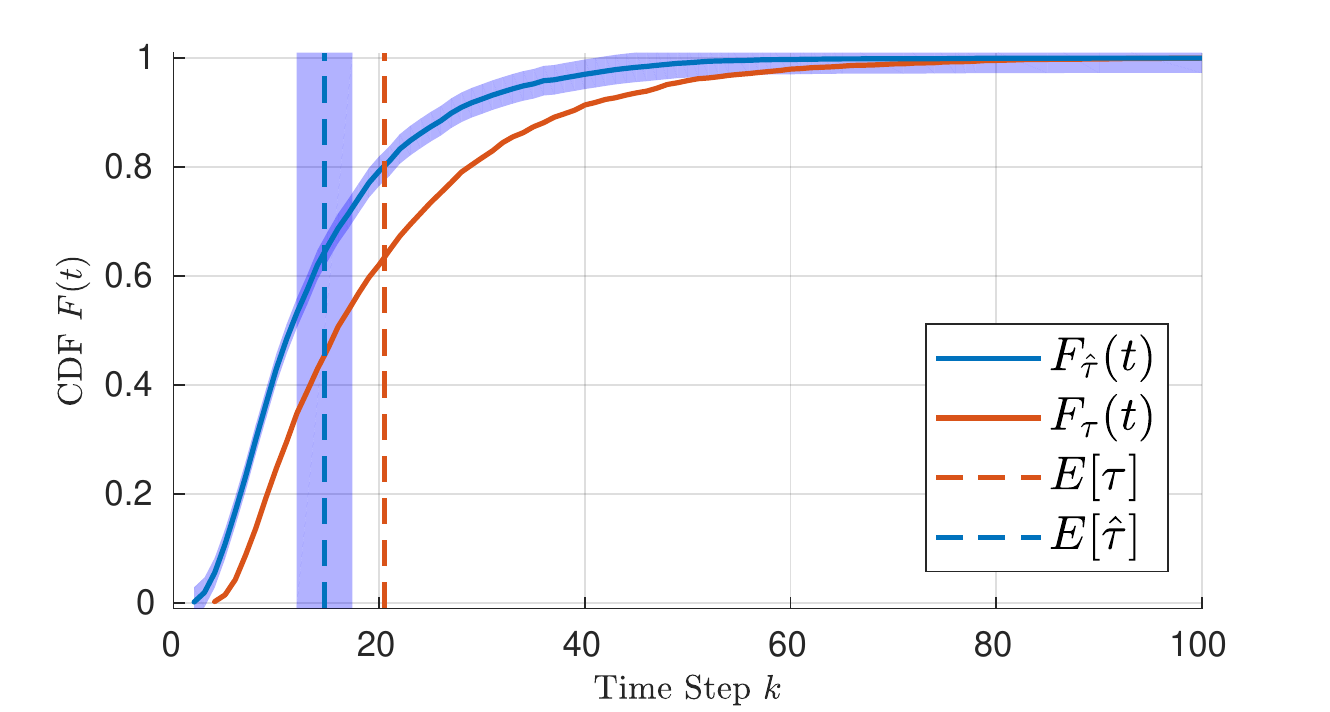}\\
    	\includegraphics[width=0.4\textwidth]{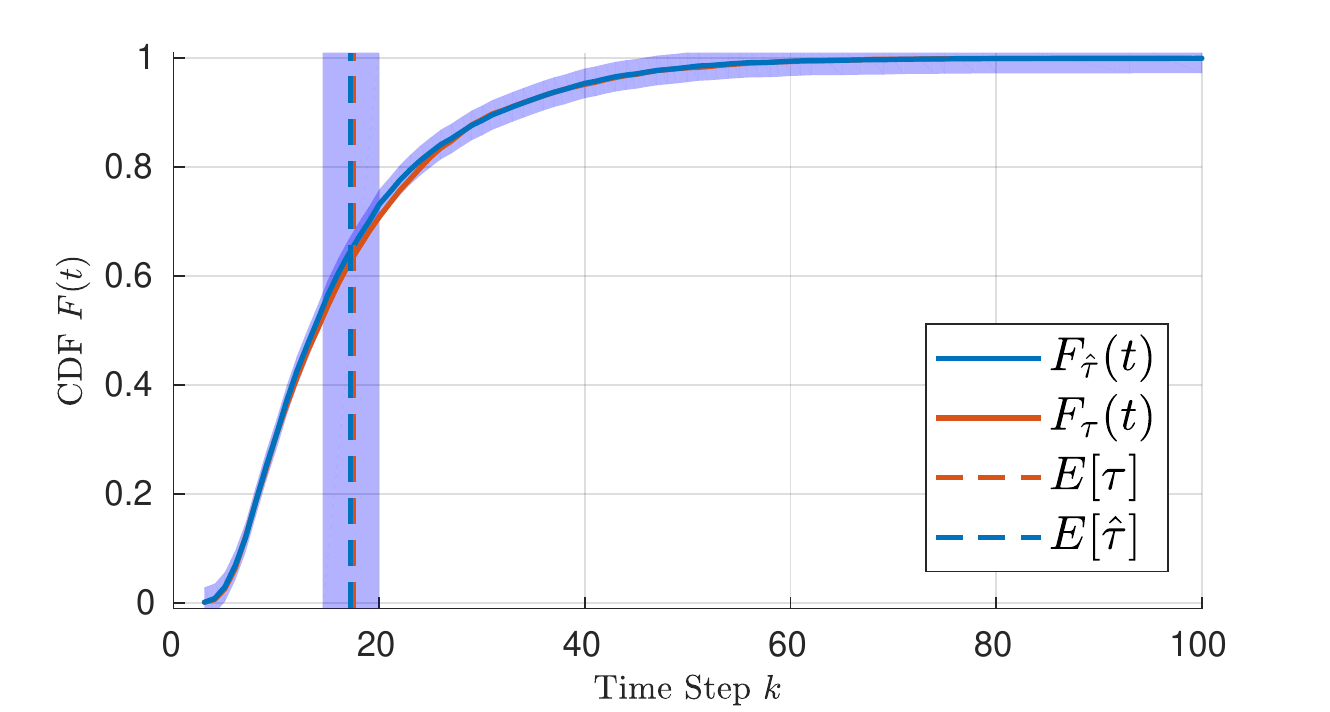}
	\end{center}
	\caption{\scriptsize Communication behavior of a system with output measurements. In the first graph, we see the test statistic for an inaccurate model and in the second for $\Hat{\theta} = \theta$. Both learning triggers (\eqref{approxtrigger} and \eqref{trigger:KStwosided}) are effective in detecting the model mismatch. Interestingly, updating the model results in more communication, as can be seen by a decrease in the actual average inter-communication time $E[\tau]$ (dashed red) between top and bottom. With the improved model, the KF tracks the true states better, and thus, we obtain more communication.}
	\label{fig:KFmorecomm}
\end{figure}
More accurate models result in better predictions.
Thus, one may expect that improved models also lead to reduced communication of state information from sender to receiver (\cf \fig \ref{fig:ETL-fullstate}).  While this is indeed the case for perfect state measurements (as has been observed in the example of \sect \ref{sec:numExample}), it may actually be the opposite for the KF setting. Here, we present an example that demonstrates this rather unexpected effect -- better models may lead to more communication.  
The reason is as follows:  better models increase the KF performance, and thus, it is possible to track the unobserved states better. Therefore, it is possible to construct examples where communication increases, which is desirable for performance, though counterintuitive.
Consider system \eqref{eq:systrue} with the matrices
\begin{equation}
\label{eq:KFmatrices}
\begin{alignedat}{1}
A = \left( \begin{array}{rrrr}
1.000 & 0.010 & -0.005 & 0.000 \\
0.017 & 1.027 & -0.301 & -0.061 \\
0.000   & 0.000  &  0.997    & 0.009 \\
0.046   & 0.067  & -0.507    & 0.850 \\
\end{array}\right) 
\end{alignedat} 
,
\begin{alignedat}{1}
C^\intercal = \left( \begin{array}{rr}
1 & 0 \\
0 & 0 \\
0 & 1 \\
0 & 0
\end{array}\right) 
\end{alignedat} 
\end{equation}
which is obtained by linearizing the closed-loop dynamics of a stabilized inverted pendulum.
We assume process noise $\epsilon(k) \sim \mathcal{N}(0,0.1 I_4)$ and observation $\nu(k) \sim \mathcal{N}(0,0.1I_2)$, where $I_n$ is the identity matrix of dimension $n$.
Further, we assume that $\hat{\nu}(k) \sim \mathcal{N}(0,0.5I_2)$ and that the model otherwise coincides with the true system parameters. We consider the KF states $\Hat{x}(k)$ (\cf \eqref{eq:sysKF}), the predictions $\check{x}(k)$, and the state trigger
\begin{equation}
\label{statetrigger:KF}
    \gamma_{\mathrm{state}}=1 \iff \| \hat{x}(k) - \check{x}(k) \|_2 \geq 1.
\end{equation}
We initialize $x(0) = \hat{x}(0) = \check{x}(0) = 0$ and obtain the distribution over stopping times depicted in \fig \ref{fig:KFmorecomm}.
The expected model-based communication is derived via Monte Carlo simulation of the innovation process \eqref{eq:KFinno}, where we set $\tau_{\mathrm{max}} = 100$ and $m=n=\num{5000}$.

\begin{figure}[tb]
	\begin{center}
		\includegraphics[width=0.4\textwidth]{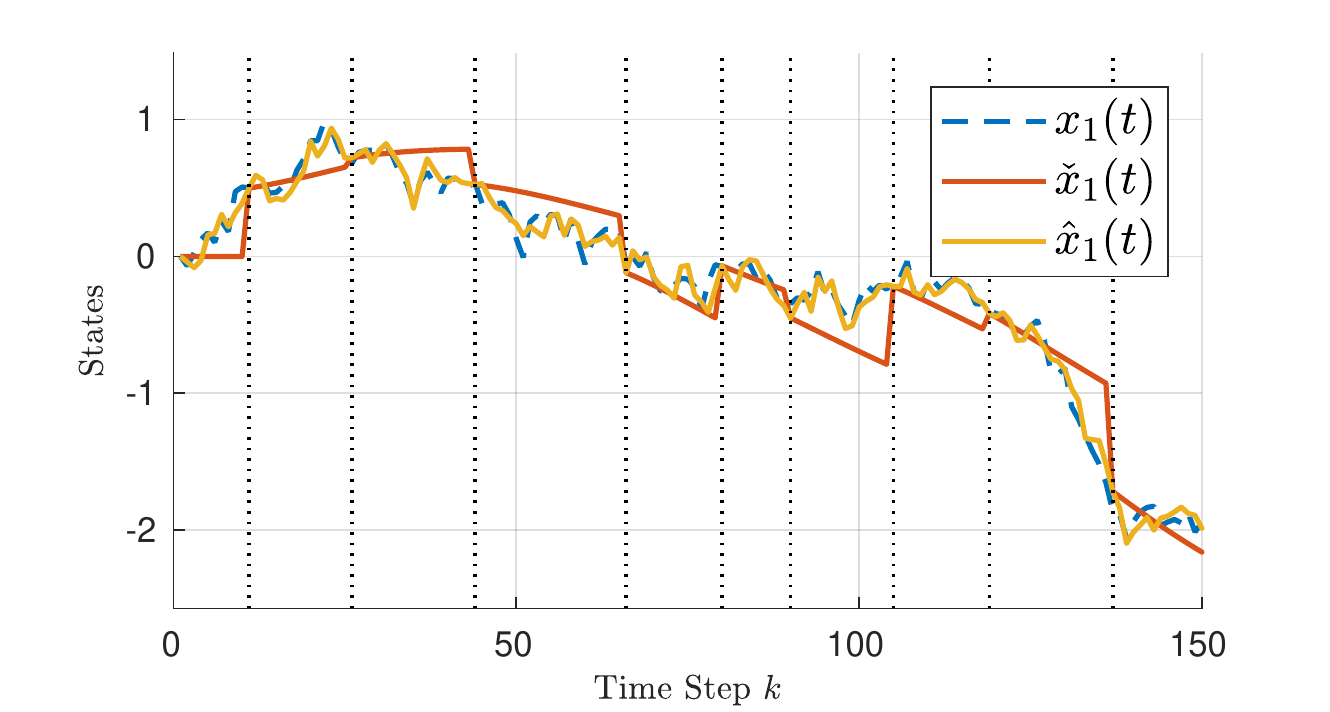}\\
    	\includegraphics[width=0.4\textwidth]{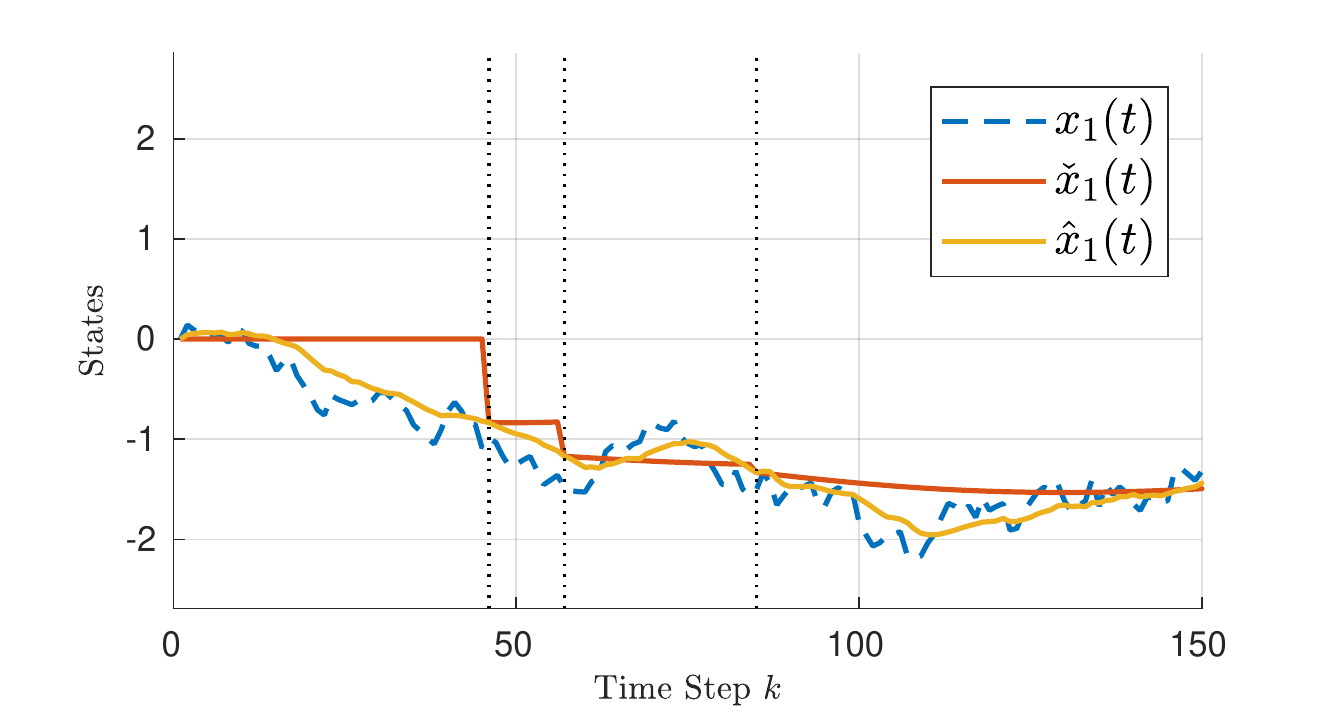}\\
    	\includegraphics[width=0.4\textwidth]{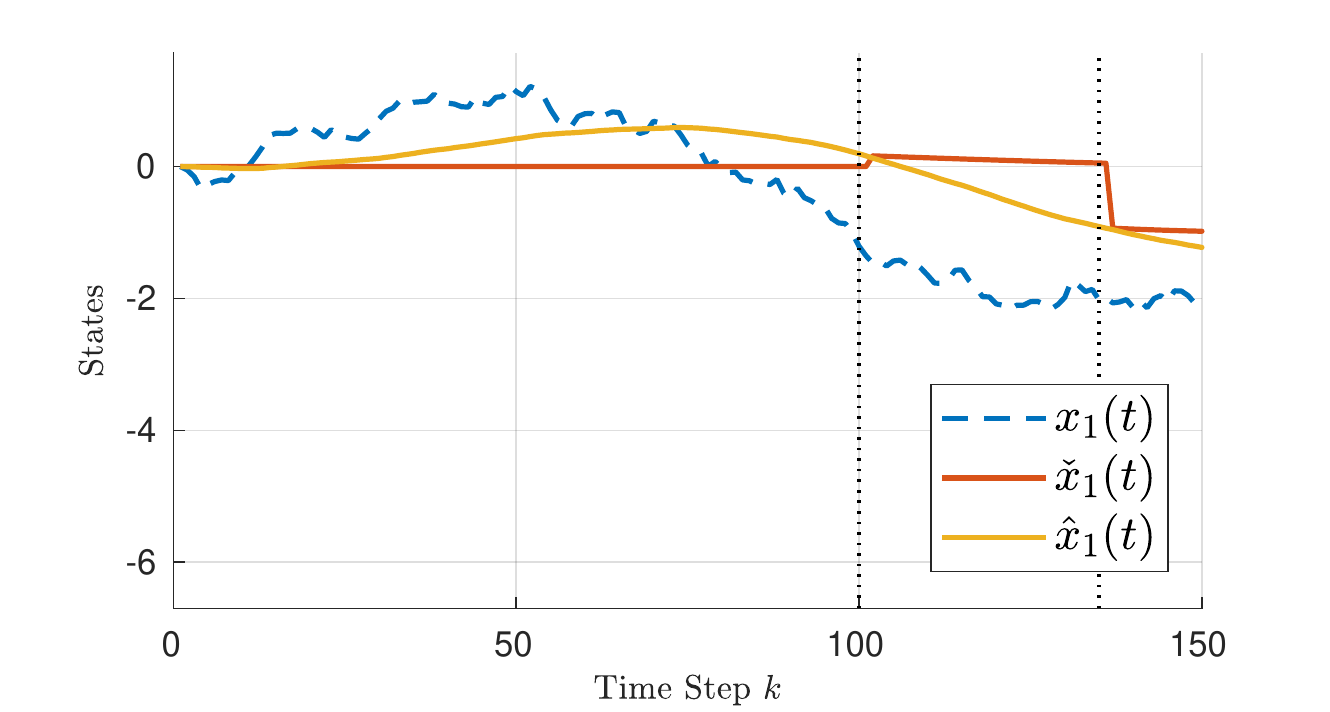}
	\end{center}
	\caption{\scriptsize State trajectories of the first dimension of the four-dimensional system with output measurements \eqref{eq:KFmatrices}. The state trigger \eqref{statetrigger:KF} is applied and therefore, communication triggered when $\Hat{x}(k)$ and $\check{x}(k)$ deviate by $\delta$. Communication instances are depicted with dotted vertical lines.
	In the first graph, we can see how well the KF $\Hat{x}(k)$ (yellow) tracks the true states $x(k)$ (dashed blue). The red line depicts the open-loop predictions $\check{x}(k)$.
	By worsening the model from the first to the second graph, the KF performance gets worse, which results in less communication. From the second to the third graph, we worsen the model even more, which results in even less communication, because the KF is not able to track the states.}
	\label{fig:KFstates}
\end{figure}

In the first graph in \fig \ref{fig:KFmorecomm}, we can see that the empirical inter-communication times are higher than the model-based. 
Updating the model actually reduces the average inter-communication time (more communication), which is because the KF improves and tracks the states $x(k)$ better, which is illustrated in \fig \ref{fig:KFstates}. The first plot shows the tracking performance when a perfect model is used $\hat{\theta} = \theta$ (KF states in yellow). For the second plot, we changed the covariance of $\nu(k)$ to $\mathcal{N}(0,0.5I_2)$, as discussed above. In the third plot, we exaggerated this effect even further by assuming $\nu(k) \sim \mathcal{N}(0,10I_2)$ in the model. In all three plots, we consider the first $k = 150$ time steps and stop afterward. In the third graph, we can see that the KF states deviate a lot from the true underlying states. However, they are still close to the open-loop predictions, and hence, there is very little communication.
Despite the counterintuitive link between model accuracy and average communication, the example shows that the derived learning triggers are effective in detecting model mismatch.

\section{Discussion and Future Work}
Event-triggered learning is proposed in this article as a novel concept to trigger model learning when needed. This article focuses on the rigorous design of learning triggers,  and we obtained (provably) effective learning triggers utilizing statistical tests.
The concept of ETL has also already been applied in hardware experiments~\citep{so18} and shown to yield reduced communication.

While event-triggered learning has been motivated as an extension to already existing methods to reduce communication in NCSs, the concept generally addresses the fundamental question of \emph{when to learn} and potentially has much broader relevance.
Here, we ultimately care about communication, and thus, it is a natural idea to analyze inter-communication times and trigger model learning based on these. Furthermore, they show advantageous statistical properties, such as being i.i.d. and scalar-valued. Depending on the concrete problem at hand, the signal used for triggering learning should be chosen accordingly. Using control performance as such a triggering signal is a potential extension, and first steps in this direction were taken in~\citep{baumann2019event}.

\bibliographystyle{agsm}        
\bibliography{biblio}           




	\vspace{2mm}
        \InsertBoxL{0}{\includegraphics[width=0.9in,height=1.1in,clip,keepaspectratio]{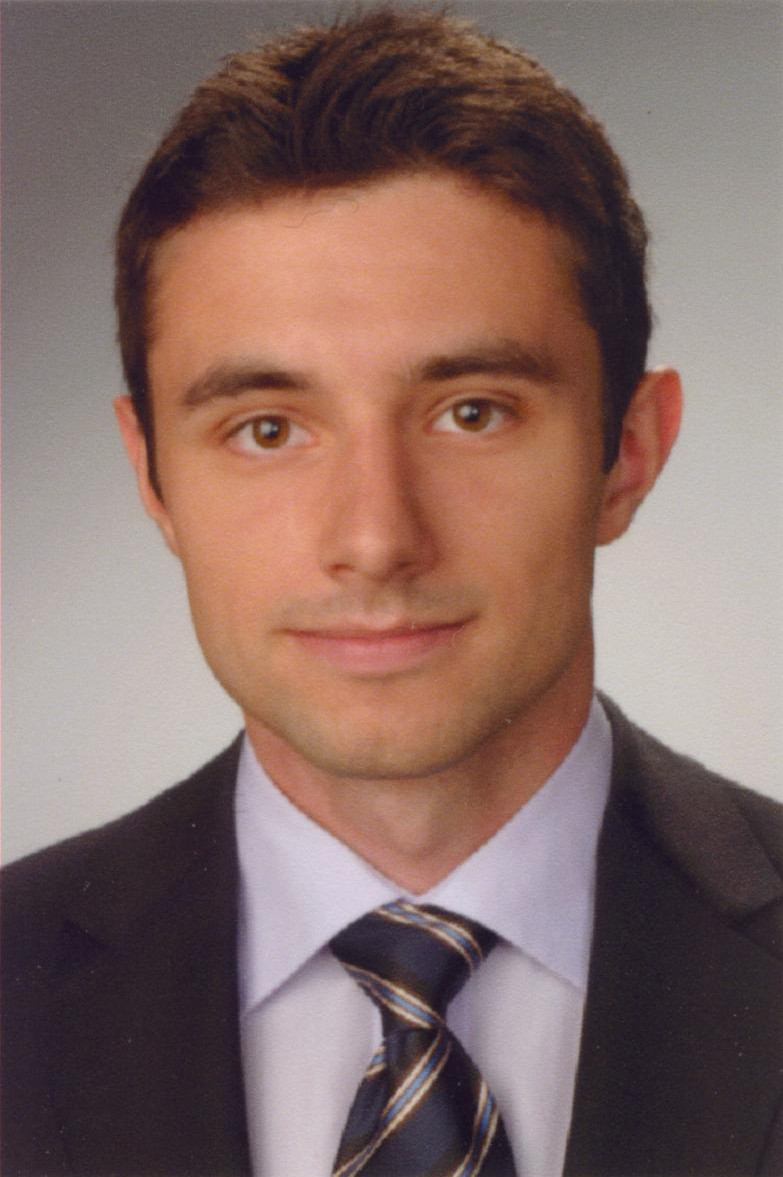}}[-1]
        \scriptsize\textbf{Friedrich Solowjow}
		%
		received B.\,Sc. degrees in Mathematics and Economics from the University of Bonn in 2014 and 2015, respectively, and a  M.\,Sc. degree in Mathematics from the University of Bonn in 2017. He  is  a  PhD  student  in the  Intelligent  Control  Systems  Group  at  the  Max Planck  Institute  for  Intelligent  Systems,  Stuttgart, Germany and a member of the International Max Planck Research School for Intelligent Systems. 
		
		\vspace{2mm}
		
		\InsertBoxL{0}{\includegraphics[width=0.9in,height=1.1in,clip,keepaspectratio]{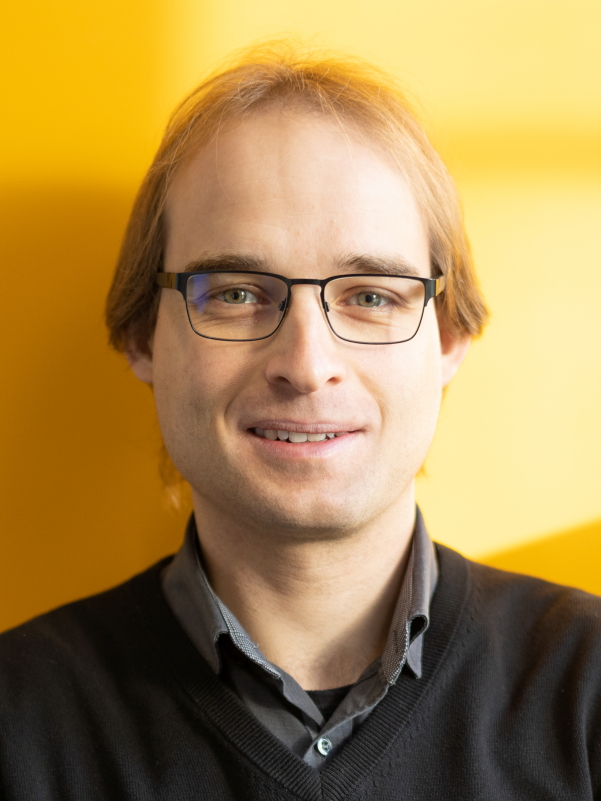}}[-1]
        \scriptsize\textbf{Sebastian Trimpe}
		%
		(M'12) received the B.\,Sc. degree in general engineering science and the M.Sc. degree (Dipl.-Ing.) in electrical engineering from Hamburg University  of  Technology,  Hamburg,  Germany,  in 2005  and  2007,  respectively,  and  the  Ph.\,D. degree (Dr.\,sc.) in mechanical engineering from ETH Zurich, Zurich, Switzerland, in 2013. He is currently   a   Research   Group   Leader   at the  Max  Planck  Institute  for  Intelligent  Systems, Stuttgart, Germany, where he leads the independent Max  Planck  Research  Group  on  Intelligent  Control Systems.  His  main  research  interests  are  in  systems  and  control  theory,machine learning, networked and autonomous systems. Dr.\,Trimpe is recipient of several awards, among others, the triennial IFAC World Congress Interactive Paper Prize (2011), the Klaus Tschira Award for achievements in public understanding of science (2014), the Best Demo Award of the International Conference on Information Processing in Sensor Networks (2019), and the Best Paper Award of the International Conference on Cyber-Physical Systems (2019).

\end{document}